\documentclass{elsart}

 \usepackage{graphicx}

\usepackage{amssymb}

\begin{document}

\begin{frontmatter}

\title{Ferromagnetism in two band metals: Combined effect of Coulomb correlation, hybridization and band widths}
\author[aff1]{C. M. Chaves}
\ead{cmch@cbpf.br}
\thanks[label1]{Partially supported by the brazilian agencies $CNP_q$ and $PCI/MCT$}
\corauth[cor1]{}
\author[aff1,aff2]{A.Troper} 
\address[aff1]{Centro Brasileiro de Pesquisas F\'\i sicas, Rua Xavier Sigaud 150,Rio de Janeiro, 22290-180, RJ, Brazil}
\address[aff2]{Universidade do Estado do Rio de Janeiro,Rua S. Francisco Xavier 524, Rio de Janeiro, Brazil}


\begin{abstract}
 We study the possibility of ferromagnetism in metals. The metal is described by two hybridized bands one of which includes Hubbard correlation whereas the other is uncorrelated. We parametrize the ratio of the band widths and their centers as well. The original Hamiltonian is transformed in an effective and simpler one. Only one site retains the full correlation (U) while in the others acts as  an  internal field, the self-energy, in the framework  of an alloy analogy approximation. This field, in turn,  is self-consistently determined by imposing the translational invariance of the problem. For several total electronic occupation numbers ($n_{total}$) we compare the spin dependent free energies with the corresponding paramagnetic ones. We present several results pointing out the mechanism by which the self-consistency introduces a sort of constraints, for given values of band width and band shift .
\end{abstract}

\end{frontmatter}

\section{Formulation of the problem}

The origin of magnetism in itinerant ferromagnets has been object of concern in last years \cite{gabi,sch,benhur,matho}. 
In an previous work \cite{sces07} we have developed a two band symmetric model to describe the effect of Coulomb correlation in these metals. It  consisted of a Hubbard like narrow band (\nolinebreak band $a$) with intrasite Coulomb interaction U,  hybridized with another band, which is broad and uncorrelated (band $b$), through the hybridization coupling $V_{ab}$. Now we allow a shift between the centers of the two bands in order to describe a more general and realistic band strucuture; we also let the band $b$ width varies. Both have strong influence on the kind of magnetic state of the metal.

As in Roth's approach\cite{roth1},  the correlation is present in only one site ( say, the origin ), while in the others acts  an effective spin and energy dependent but site independent field, the self-energy  $\Sigma^{\sigma}$. This field is self-consistently determined by imposing the vanishing of the scattering $T$ matrix associated to the origin, thus restoring the translational invariance of the host. 
The effective Hamiltonian is 
\begin{eqnarray}
\label{Heff}
\mathcal{H}_{eff}&=&\sum_{i,j,\sigma }t_{ij}^{a}a_{i\sigma }^{+}a_{j\sigma
}+\sum_{i,j,\sigma }t_{ij}^{b}b_{i\sigma }^{+}b_{j\sigma
}\\
&+&\sum_{i \not=0,\sigma}n_{i\sigma}^{a}\Sigma^{\sigma}+Un^{a}_{0\uparrow }n^{a}_{0\downarrow} 
+\sum_{i,j,\sigma }(V_{ab}b_{i\sigma }^{+}a_{j\sigma }+ h.c.) 
\end{eqnarray}
where  $n_{i\sigma }^{a}=a_{i\sigma }^{+}a_{i\sigma }$ and $\sigma$ denotes spin. $t_{ij}$ denotes the tunneling amplitudes between neighboring sites $i$ and $j$ , in each band.
Even with this single-site approximation (SSA), $\mathcal{H}_{eff}$ describes in a quite satisfactory way the  physics of these systems but we still need to resort to further approximations. And these include the decoupling in the Green function  and the self consistency generated by the vanishing of the scattering $T$ matrix. The main equations we get are:
\begin{equation}
G^{a}_{kk',\sigma}(w)={\frac{\delta_{kk'}}{w-\tilde{\epsilon}^a_k -\Sigma^\sigma(w)}}, 
\label{gd}
\end{equation}
the Green function of the  $a$ band,  
%
%
%
%
where 
\begin{equation}
\tilde {\epsilon}^{a}_{k}=\epsilon^{a}_k +{\frac{|V_{ab}|^2(k)}{w-{\epsilon}^b_k }}, 
\label{etild}
\end{equation}
is the recursion relation of the renormalized $a$ band and $\epsilon^{a}_k$ and $\epsilon^{b}_k$ denote the bare bands, with
\begin{equation}
\epsilon^{a}_{k}={\frac{t_{a}(cos(k_xa)+\cos(k_ya)+\cos(k_za))}{A}}. 
\label{ed}
\end{equation}
In this paper we use $t_a =1$ and $A=3$, in arbitrary energy units. All energy magnitudes are taken in units of $t_a$, making them dimensionless. The bare $a$ band width is then $W=2$. 

We use homothetic bands
\begin{equation}
\label{homo}
\epsilon^{b}_{k}=\epsilon_s +\alpha \epsilon^{a}_{k}.
\end{equation}
$\epsilon_s$ is the center of the $b$ band and represents a shift in the bands. $\alpha$ is a phenomenological parameter describing the ratio of the effective masses of the $a$ and the $b$ electrons. From now on we take  $k_ia \rightarrow k_{i}, i=x,y,z$ and $V_{ab}=V_{ba}\equiv V=$ real and constant independent of $k_i$.

\begin{figure}[!ht]
\label{bare}
\begin{center}
\includegraphics[angle=0,width=0.45\textwidth]{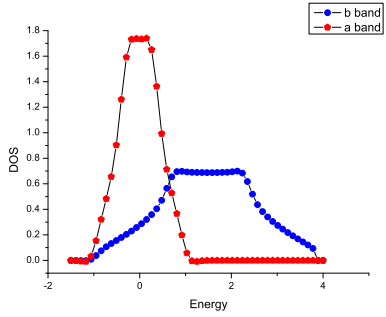}
\end{center}
\caption{DOS of the bare $a$ and $b$ bands ($U=0$, $V=0$) for $\alpha = 2.5$ and $\epsilon_s=1.5$.}
\end{figure}

In Fig. \ref{bare} we display  typical value for the density of states (DOS) of the bare bands when $\alpha=2.5$ and $\epsilon_s=1.5$.

The vanishing of the T-matrix gives a self-consistent equation for the self-energy:
\begin{equation}
\label{elivre}
\Sigma^\sigma= U\langle n^a_{0-\sigma}\rangle+(U-\Sigma^\sigma)F^\sigma(w,\Sigma^\sigma)\Sigma^\sigma,
\end{equation}
with
\begin{equation}
\label{F}
F^\sigma(w,\Sigma^\sigma)= N^{-1}\sum_k G^{a}_{kk,\sigma},
\end{equation}
where  $\langle n^a_{0-\sigma}\rangle$ is an initial guess for $a$-band occupation number at the origin.

\section{The method in action : numerical results}

The self-consistency is performed both in $\Sigma^\sigma$ and in $\langle \nolinebreak n_{0,\sigma}^a\rangle$, for each total occupation $n_{total}=\langle n^a \rangle +\langle n^b \rangle $. The total number of electrons per site is a constant of motion. The self-consistency and the dynamics distribute the electrons among $\langle n_{0,\uparrow}^a\rangle$, $\langle n_{0,\downarrow}^a\rangle$ and $\langle n^b \rangle$, eventually producing a ferromagnetic state (it turns out that the $b$-band is allways paramagnetic for the range of parameters used in this work). We want now to exhibit the combined effect of $U$, $V$, $\alpha$ and $\epsilon_s$ on this process at $T=0K$. The effect of $V$ and $\epsilon_s$ , apart from transfering electrons from one band to the other, is to produce band broadening\cite{sch} in both bands.

\begin{figure}[!ht]
\begin{center}
\includegraphics[angle=0,width=0.45\textwidth]{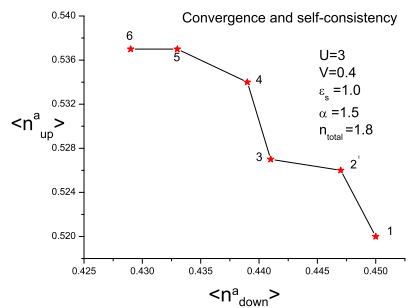}
\end{center}
\caption{This ilustrate the process of convergence through self-consistency for the $\langle n^{a}_{\uparrow}\rangle$ and $\langle n^{b}_{\downarrow}\rangle$. Point $1$ was our initial guess for $U=3$, $V=0.4$, $\alpha=1.5$, $\epsilon_s=1.0$ and $n_{total}=1.8$. Then the path $1\rightarrow 2\rightarrow 3\rightarrow 4\rightarrow 5\rightarrow 6$ is generated in the process.  Only in the last step the coordinates of the output, point  $6$, differ from that of the previous point, $5$, by less than $eps=0.005$.} 
\label{conv}
\end{figure}

In Fig. \ref{conv} it can be seen the process by which an initial guess, denoted by the point  $(\langle n_{0,\uparrow}^a \rangle,  \langle n_{0,\downarrow}^a \rangle) $, labeled $1$, converges, here after five steps,  for $U=3$, $V=0.4$, $\alpha=1.5$ and $\epsilon_s = 1.0$, to point $6$, whose coordinates differ from those of point $5$ by less than $eps=0.005$. In some situations the number of steps needed to achieving convergence may be larger.

\begin{figure}[!ht]
\begin{center}
\includegraphics[angle=0,width=0.45\textwidth]{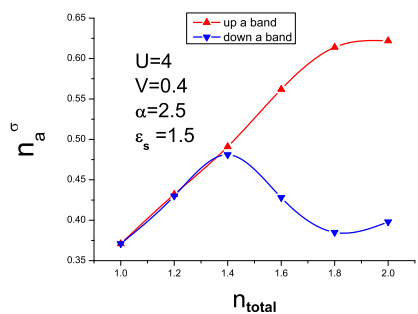}
\end{center}
\caption{Converged values of $\langle n^a_{\sigma} \rangle$ for an interval of $n_{total}$ for $U=4$, $V=0.4$, $\alpha=2.5$ and $\epsilon_s=1.5$.}
\label{nd2}
\end{figure}

In Fig. \ref{nd2} we plot the converged values of $\langle n^a_{\sigma}\rangle $,$\sigma= \uparrow, \downarrow$ in an interval of  $n_{total}$ for $U=4$, $V=0.4$, $\alpha=2.5$ and $\epsilon_s=1.5$. For $n_{total}$ less than $\sim 1.4$  the system is basically paramagnetic.

\begin{figure}[!ht]
\begin{center}
\includegraphics[angle=0,width=0.45\textwidth]{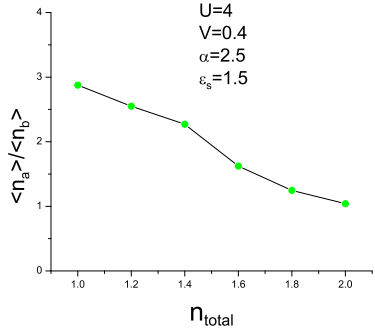}
\end{center}
\caption{Ratio of the converged values of $\langle n^a \rangle / \langle n^b \rangle$ for an interval of $n_{total}$. Here $U=4$, $V=0.4$, $\alpha =2.5$ and $\epsilon_s =1.5$. It is seen that for this values of the parameters $\langle n^a \rangle \gg \langle n^b \rangle$.}
\label{nanb}
\end{figure}

In Fig. \ref{nanb} we exhibit the ratio of $n_a/n_b$ versus $n_{total}$ as obtained from the process of convergence and self-consistency. Here $U=4$, $V=0.4$, $\alpha=2.5$ and $\epsilon_s=1.5$. It is seen that for small $n_{total}$ the method distributes most of the electrons into the $a$-band.

\begin{figure}[!ht]
\begin{center}
\includegraphics[angle=0,width=0.45\textwidth]{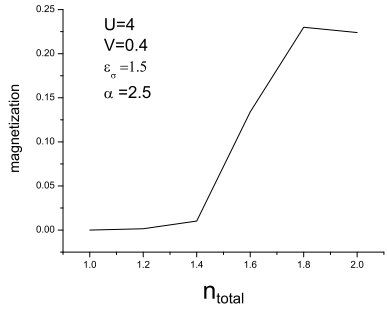}
\end{center}
\caption{$a$-band magnetization versus $n_{total}$ for $U=4$, $V=0.4$, $\alpha=2.5$ and $\epsilon_s=1.5$}.
\label{mag}
\end{figure}
\begin{figure}[!ht]
\begin{center}
\includegraphics[angle=0,width=0.45\textwidth]{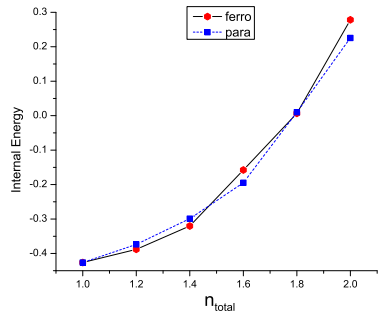}
\end{center}
\caption{Free energy of the metal versus $n_{total}$ for $U=4$, $V=0.4$, $\alpha=2.5$ and $\epsilon_s=1.5$ for the paramagnetic  and the ferromagnetic states.}
\label{Elivre}
\end{figure}
\begin{figure}[!ht]
\begin{center}
\includegraphics[angle=0,width=0.45\textwidth]{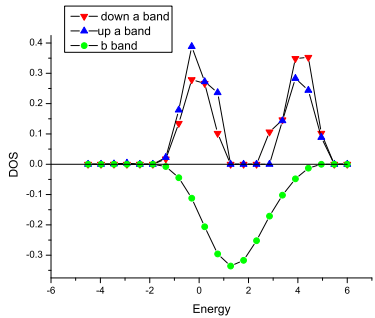}
\end{center}
\caption{DOS of the $a$ and $b$ bands. The latter is non magnetic. Here $U=4$, $V=0.4$, $\alpha=2.5$, $\epsilon_s=1.5$ and $n_{total}=1.6$. The Fermi energy is at $E_F =0.872$ and the magnetization is $m=0.134$. Both bands broad under the effct of $V$ and $\epsilon_s$.}
\label{dosab}
\end{figure}

In Fig. \ref{mag} it is plotted magnetization versus $n_{total}$ when $U=4$, $V=0.4$, $\alpha=2.5$ and $\epsilon_s=1.5$. In agreement with Fig. \ref{nd2} the magnetization starts at $n_{total}\sim 1.4$ and increases monotonically up to $n_{total}=1.8$. For the same values of parameters the internal energy of the ferromagnetic state is compared with that of the paramagnetic one in Fig. \ref{Elivre}, for the same range of $n_{total}$.
The difference between them is rather small and oscilatting with $n_{total}$. Several factors influence on this: the small separation of the correlated $a^{\sigma}$ bands and the contibution of the uncorrelated $b$ band, which may be of the same order in both states and even overcomes the contribution of the $a$ band for some occupation $n_{total}$. From Fig. \ref{dosab} it can be seen that the $b$ band contribution competes with  the one from the $a$ band if $E_F$ crosses the Hubbard gap. So we expect that band degeneracy may be crucial to stabilize the ferromagnetic state.
A systematic study of this problem, varying the paramenters $U$, $V$, $\alpha$ , $\epsilon_s$, including the degeneracy of the $a$ band that allows also a exchange term, as well the generation of a phase diagram is now in progress\cite{ch}.

\nopagebreak

\end{document}